\documentclass[11pt]{article}
\makeatletter\if@twocolumn\PassOptionsToPackage{switch}{lineno}\else\fi\makeatother

\usepackage{graphicx}

\usepackage[T1]{fontenc}

\usepackage{lineno}

\usepackage{natbib}

\usepackage{JASA_manu}

\usepackage{tabulary,amsmath,amssymb,amsfonts,amsbsy,xcolor}

\usepackage{url,multirow,morefloats,floatflt,cancel,tfrupee, float}
\makeatletter

\AtBeginDocument{\@ifpackageloaded{textcomp}{}{\usepackage{textcomp}}}
\makeatother
\usepackage{colortbl}
\usepackage{xcolor}
\usepackage{pifont}
\usepackage[nointegrals]{wasysym}
\makeatletter

\def\BreakURLText#1{\@tfor\brk@tempa:=#1\do{\brk@tempa\hskip0pt}}
\let\lt=<
\let\gt=>
\def\processVert{\ifmmode|\else\textbar\fi}

\makeatletter\def\oupIndent{1pt}
\def\author#1{\gdef\@author{\hskip-\dimexpr(\tabcolsep)\hskip\oupIndent\parbox{\dimexpr\textwidth-\oupIndent}{\centering#1}}}

\makeatother

\begin{document}

\title{Best Practices for Collecting Gender and Sex Data\\ Word Count: 6027}
\author{Suzanne Thornton\thanks{Corresponding author. E-mail: sthornt1@swarthmore.edu}{ }\\Assistant Professor\\Mathematics and Statistics Department\unskip, Swarthmore College\\500 College Ave, Swarthmore, 19081, PA, United States\\email: \texttt{sthornt1@swarthmore.edu}
\\Dooti Roy\\Principle Methodology Statistician\unskip, Boehringer Ingelheim\\Danbury, 06810, CT\\email: \texttt{dootiroy@gmail.com}
\\Stephen Parry\\Statistical Consultant\unskip, Cornell University\\Ithica, 14850, NY, United States\\email: \texttt{sp2332@cornell.edu}
\\Donna LaLonde\\Director of Strategic Initiatives and Outreach\unskip, American Statistical Association\\Alexandria, 22314, VA, United States\\email: \texttt{DonnaL@amstat.org }
\\Wendy Martinez\\Director, Mathematical Statistics Research\unskip, Bureau of Labor Statistics\\Washington, DC, 20212, United States\\email: \texttt{martinez.wendy@bls.gov }
\\Renee Ellis\\Family Demographer\unskip, US Census Bureau\\Washington, DC, 20002, United States\\email: \texttt{reneerellis@gmail.com}
\\David Corliss\\Lead data scientist, manager of the Data Science Center of Excellence\unskip, Stellantis \\Auburn Hills, 48309, MI, United States.\\email: \texttt{davidjcorliss@gmail.com}}

\maketitle 

\newpage \begin{center} \textbf{Abstract} \end{center}
The measurement and analysis of human sex and gender is a nuanced problem with many overlapping considerations including statistical bias, data privacy, and the ethical treatment of study subjects. Traditionally, human gender and sex have been categorized and measured with respect to an artificial binary system. The continuation of this tradition persists mainly because it is easy to replication and not, as we argue, because it produces the most valuable scientific information. Sex and gender identity data is crucial for many applications of statistical analysis and many modern scientists acknowledge the limitations of the current system. However, discrimination against sex and gender minorities poses very real privacy concerns when collecting and distributing gender and sex data. As such, extra thoughtfulness and care is essential to design safe and informative scientific studies. In this paper, we present statistically informed recommendations for the data collection and analysis of human subjects that not only respect each individual's identity and protect their privacy, but also establish standards for collecting higher quality data.

\vspace*{.3in}\def\keywordstitle{Social construct, categorical data, identifiability, gender and sex minorities.}
    
\section{Introduction}
When teaching introductory statistics, many instructors encounter the question: Why do we typically use a significance level of 0.05 (or 0.01 or 0.10)? The answer is simple but not very gratifying: this is a matter of convention. Helpful guidelines are well-established but there are infinite possible significance levels a researcher could decide to use. A similar conundrum occurs when collecting data on socially-defined attributes such as race or gender. The typical ways of categorizing these variables have been passed down from statisticians to medical practitioners and other researchers more often due to their ease of replication rather than their actual scientific value. However, qualities such as these represent social constructs which, by their very nature, are not fixed but are fluid concepts with evolving definitions. Merriam-Webster defines the term social construct as ``an idea that has been created and accepted by the people in a society''. Scientific endeavors must recognize that many qualitative human characteristics fit this definition and as researchers, we are responsible for practicing extra thoughtfulness and care when considering these traits as variables of interest in our studies. Ultimately, we cannot adequately provide scientific or medical benefits to socially marginalized people if we are not collecting the right data. 

The goal of this document is to present statistical practitioners with guidelines on how to collect quality data on sex and gender for adult subjects. This work is meant to serve as an introduction to data quality concerns and ethical concerns from a perspective of gender inclusivity, a perspective that has not traditionally been emphasized in the history of data science. We believe it is time to challenge convention and strive for better practices when collecting and analyzing human-centric data. This document is not a definitive source for understanding the complexities of gender identity but does provide an introduction to the nuances of gendered language from a statistical perspective and outlines a framework for researchers to adopt to better devise useful to address this specific challenge in quantifying the human experience. These recommendations are meant to be revisited and updated as our society evolves. 
    
\section{Background information and terminology}
While a thorough introduction to the language of sexual and gender identities is beyond the scope of this paper, we have listed some terms and definitions in Figure 1 which will be referenced throughout the remainder of the text. Many other reputable resources elaborate upon these terms and present more terminology than we do, e.g. ~\citet{BatesEtAl2019}; ~\citet{BauerEtAl2019}; and ~\citet{Badgett2009}. For additional resources, see Appendix A.

\begin{figure}[H]
%\begin{center}
\includegraphics[width=\linewidth]{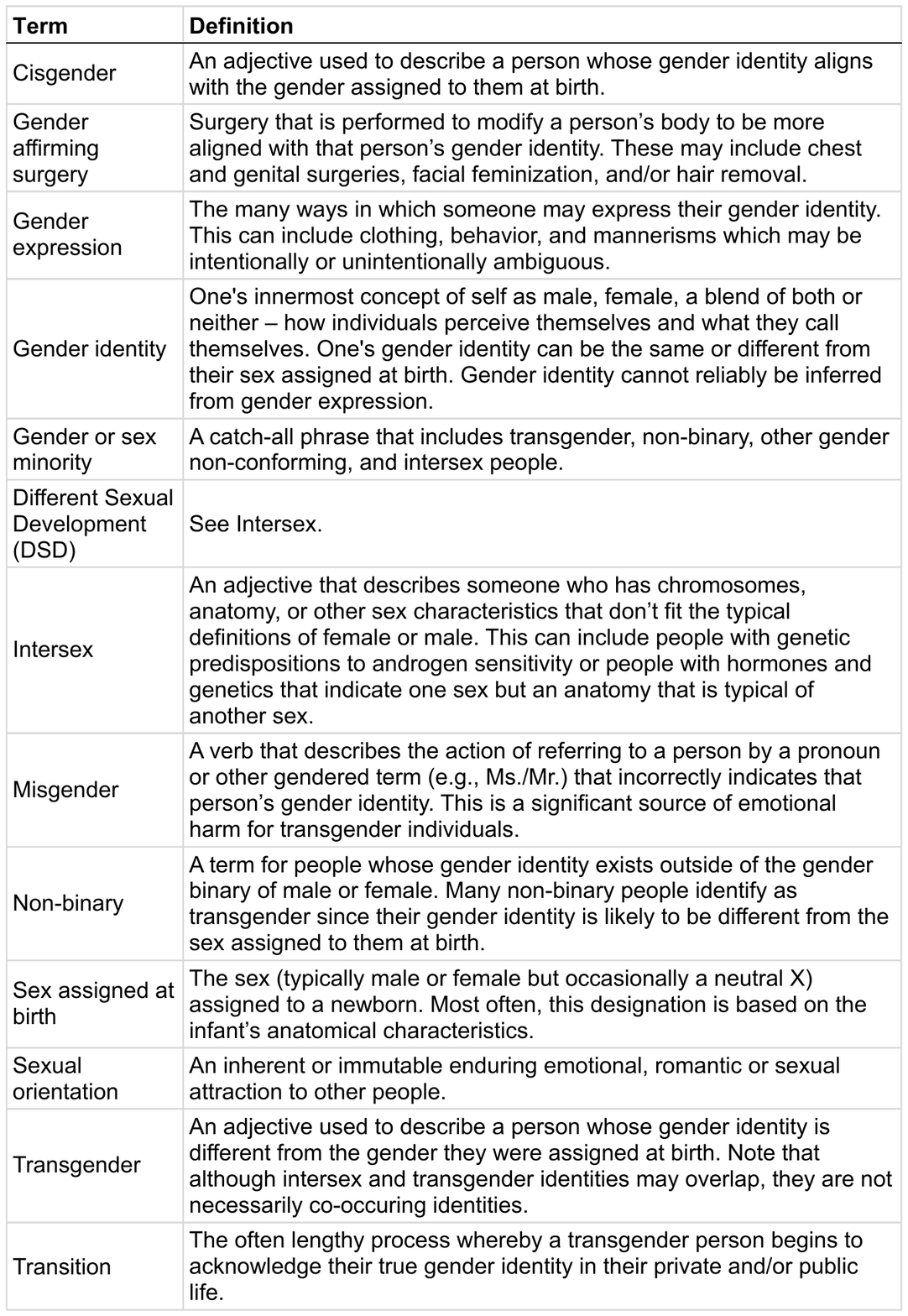}
%\end{center}
\caption{Common terms relevant to gender and sex minorities. Definitions are derived from the National LGBTQIA+ Health Education Center and the Human Rights Campaign.}
\end{figure}

It is important to note the many possible intersections of LGBTQ+ identities. Naturally, the full scope of people's identities extends beyond any one particular categorization. Human beings exist as unique intersections of multiple identities, some of which are ascribed by others and some of which are personal and may not be readily identifiable to others. Figure 2 presents a simple diagram that illustrates the possible (in)congruencies among gender and sex characteristics.

\begin{figure}[h]
\begin{center}
\includegraphics[width=0.7\textwidth]{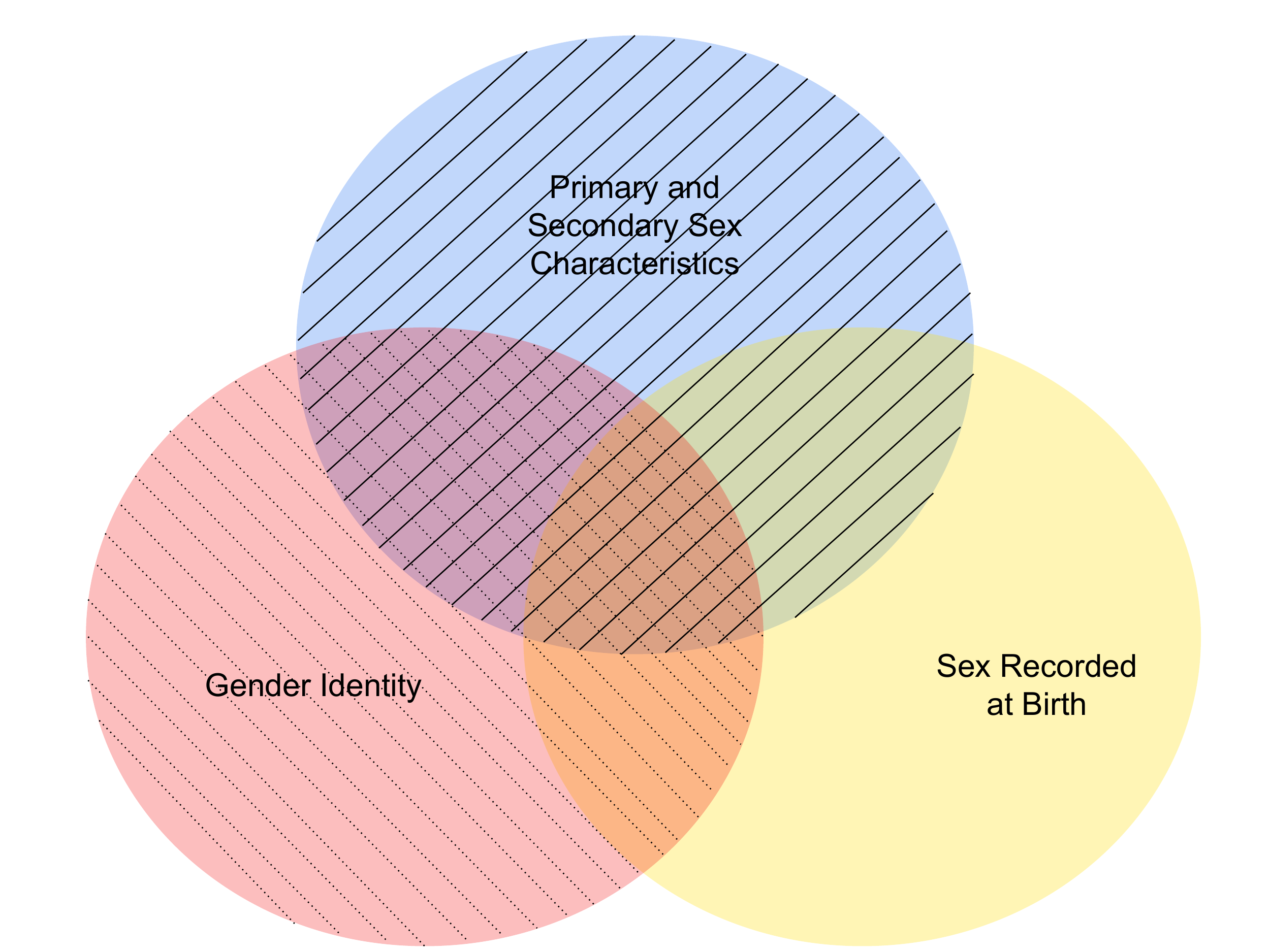}
\caption{A Venn Diagram show three different aspects of sex and gender identity. For many people, all three characteristics align; however, for gender and sex minorities one or more of these characteristics may not align in the typical manner.}
\end{center}
\end{figure}

\subsection{Sexuality}We are not discussing sexuality in this paper but we mention it here because sexual preference is often confused with the concepts of gender identity and sex. While sex and gender are descriptors of one's self and personal identity, sexuality is a descriptor of one's romantic attraction to others. 

For example, a transgender woman may identify as straight if she is only romantically interested in men; a cisgender man may identify as gay if he is attracted to other men who may or may not be cisgender. It is important for data scientists to be aware of this distinction because potential biases can be introduced in a data set by focusing on one identity but using general, catch-all terms like ``LGBTQ+'' or ``queer'' {\textemdash} neither of which specify identify gender identity or sexuality.

\subsection{Gender}Though a nuanced understanding of gender may seem like a new trend to some, the modern recognition of gender as a continuum actually has roots that stretch across the globe and into the ancient past; c.f. e.g. ~\citet{WiesnerHanks2011} and~\citet{Herdt2020}. For modern scientists to be advocating for an understanding of gender identity that extends beyond a binary classification system, is actually an example where modern science is only just now catching up to global historical and cultural wisdom. 

The enornaminity of gender identity is an immutable factor of the human experience; the language with which we describe that experience however is non-constant and changes with generations. For example, although the Chicago Manual of Style renounced the formal use of ``they'' as a generic third-person singular pronoun in 2017, two years later the American Psychological Association endorsed the formal use of ``they'' to refer to a person whose gender identity is unknown or irrelevant to the context of the usage. Regardless of the social norms of the times, the purpose of data collection has always been to understand a particular population by capturing information in bite-sized, digestible pieces. Although information loss is an inevitable byproduct of this process when applied to social constructs such as gender, our goal as statistical practitioners is to design surveys and experiments to capture the most relevant bits of information as possible. 

The concept of gender in modern society can be broken into two components, gender expression and gender identity. Respecting someone's gender identity is an ethical practice that honors the individual's sense of self. Gender identity is a personal understanding of oneself; it comes from within, not without. Gender expression refers to many possible expressions of gender including attributes such as clothing, hair, mannerisms, postures, etc. Gender expression is like art in that it may be formed by an artist who wants to communicate something particular but is subject to the interpretation of the viewers who may or may not understand the artist's intent. For any individual, gender identity and gender expression can change with time. For example, a transgender woman may not have transitioned until later in her life and thus in her younger years may have adopted masculine or androgynous gender expression for various reasons including her personal safety and comfort. The process of transitioning in one's public and private life takes years and can be a non-linear process. Some transgender men for example, may identify as non-binary as they begin to question the gender identity imposed upon them from birth and then realize their identity as men years later. Gender expression can be intentionally ambiguous or it can challenge the status quo. For all of these reasons, it is best to never assume any individual's gender identity from their gender expression alone.

A common way to communicate gender identity is to share one's pronouns, similar to sharing a surname or a title. Asking someone which pronouns they use and providing your own is a way to actively respect diverse gender identities. Pronouns can indicate information regarding how one identifies themselves along the gender spectrum, from masculine to feminine and beyond. Some people use gendered pronouns such as she/her/hers or he/him/his, other people may only use the gender neutral singular they/them/theirs, and others still may use different gender-neutral pronouns such as ze/zim/zir. In line with the American Psychological Association, we recommend the usage of the singular ``they'' as an appropriate and respectful way to refer to people whose gender identity is unconfirmed or non-binary.

\subsection{Sex}Despite the overwhelming tendency to treat biological sex as a binary variable, the term ``biological sex'' is not quite so cut and dry. When sexual characteristics of individuals are a variable of interest in a study this can refer to primary or secondary sexual attributes, genetic information, gender assigned at birth, or many other sexual characteristics. The most common use of the phrase ``biological sex'' seems to refer either to a person's chromosomes or to their genitalia at birth, however, neither of these attributes are actually binary~\citep{Ainsworth2018}.

A consensus statement published in 2005 endorsed the terminology disorded/different sex development (DSD) to describe individuals with congenital conditions in which the development of chromosomal, gonadal, or anatomic sex is atypical. However, the term intersex is preferred by many activists and advocates since it is not affiliated with a derogative connotation like the word ``disorder'' which implies something that needs to be ``fixed''~\citep{HughesEtAl2006,interACT}. In this article we will therefore refer to anyone with these congenital conditions as intersex. 

Intersex people can be born with the differences mentioned above, or they may develop these differences later in childhood. This makes pinning down the incidence rate very difficult. Most experts agree that the prevalence of intersex births is about 2\% of the population which, for comparison, is similar to the prevalence of red hair; c.f. e.g. ~\citet{Onyiriuka2016} or ~\citet{FaustoSterling2000}. People with ambiguous or atypical sexual characteristics are often coerced into changing their bodies, usually at a very young age. Most surgeries to change intersex traits happen in infancy and it is possible that an adult may not be aware of this procedure in their medical history as noted by~\citet{GreenbergStam2012}.

We bring up these points to illustrate that when one is conducting a study that merely asks participants for their sex, one risks losing important information by making the assumption that all participants understand ``biological sex'' in the same way. This is true not just for intersex participants but also can apply to transgender patients who may have undergone gender confirming genital reconstructive surgery. When collecting data on attributes that can have different meanings, it is best for the researcher to be as explicit as possible about their working definitions to avoid misclassification bias.
    
\section{A multifaceted argument for inclusive data}
Among the many reasons why researchers may want to be more inclusive in their data collection process, here we focus on two: the ethical treatment of human subjects and the collection of better quality data. While we do not claim to be experts in ethics, data is our specialty, and we have found that these two arguments overlap. Any ethically responsible researcher or clinician collecting data on human participants should be aware of the language needed to correctly identify different segments of the population. When options outside of the gender binary are not provided in surveys, for example, the researcher overlooks an entire group of people from their analysis. Or when individuals do not see their identities respected in the data collection process, they may experience feelings of confusion, depression, or failure as a result. On a larger scale, exclusion of gender and sex minorities means that research results may be unable to identify smaller sub-populations who may be the most vulnerable to certain illnesses. 

For instance, previous research has shown that transgender individuals might be at a higher risk than the cisgender population for a variety of health issues such as stress, blood pressure, and heart attacks~\citep{AlzahraniEtAl2019,Dubois2012,Irwig2018}. Mental health issues are also disproportionately prevalent among gender and sex minorities, and, as is widely acknowledged within the LGBTQ+ community, the main factors contributing to these issues stem from perceived and realized threats to personal safety and social inclusion; c.f. e.g.~\citet{OlsonMCLaughlin2016}; ~\citet{ClementsNolle2006}; or~\citet{PascoeRichman2009}. By collecting relevant information in a way that respects gender identities, researchers can help mitigate the crippling barriers to health care than many transgender and non-binary people face. For example, a 2018 study found that transgender people whose pronouns were respected by all or most people in their lives attempted suicide at half the rate of those who did not have their pronouns respected~\citep{RussellEtAl2018}. 

The fundamental goal of statistics is to articulate information about some population through the collection and analysis of data. The ethical statistician must use methodology and data that are relevant and appropriate, without favoritism or prejudice, in a manner that produces valid, interpretable, and reproducible conclusions~\citep{ASAethics}. There is much information to be learned from gender and sex qualities of human subjects provided these qualities are researched in a way that respects the variability of identities and experience. 

As is true for data collection at large, when gender and sex are attributes collected without forethought, the data can be easily and irreparably corrupted by statistical biases. A study is susceptible to non-response bias, for example, when survey participants do not see their identities represented in the questions or answers. In such situations, participants may become disgruntled and not complete the survey and therefore the conclusions from statistical analyses of the data may not apply to important subgroups of the population. Of course misclassification bias is also possible when surveys are not designed to acknowledge different identities. If a clinician assumes a participant's gender identity based on their appearance, voice, or mannerisms, the information encoded in the survey may be irrelevant, or worse, inaccurate.

With both response and non-response bias, the observed data fail to accurately represent the population of interest. Ultimately this can lead to a loss in the signal of the response variable. Although it will take time and effort to transition to more inclusive data collection practices for gender and sex, these efforts will eventually result in better quality data and more useful analyses. With more informative data, signal detection may become easier and resulting statistical models may be useful for a wider range of the population. Example 1 considers a setting in which these biases can be hazardous to the health of transgender patients in particular. 

\textbf{Example 1}

\textit{\textbf{Setting:} Healthcare providers often hesitate to prescribe drugs citing that there is no information about the safety of the drug on the specific population on the drug label. If there are no data on a specific group of people and the healthcare providers suspect that the subgroup population may react differently than the rest, they may be reluctant to prescribe the drug. A number of studies have highlighted the increased risk of mental illnesses (e.g. depression, anxiety, and suicidality) in gender and sex minorities, e.g.} \textit{~\citet{SuEtAl2016}. For clinical trials that are focused on treatments for mental illnesses, it is therefore crucial to identify individuals who may have a higher risk of certain adverse events. Furthermore, transgender and non-binary patients may take large doses of hormones as part of their transition and hormone therapy may significantly influence and impact the safety or efficacy of a particular drug. For these reasons, it is especially relevant for these trials to be able to identify transgender patients to ensure that a drug is safe and effective across all populations.}

\textbf{\textit{Problem:}} \textit{ Most modern clinical trials are not designed to identifying transgender patients, even if the patient voluntarily self-identifies. Also, within a clinical trial setting, researchers are required to justify and explain why certain information is collected and needed for the study.}

\textbf{\textit{Proposed solution:}} \textit{ Collect both ``sex assigned at birth'' and ``gender identity'' as the baseline characteristics during the conduct of the clinical trial. If these two fields do not align, this indicates a patient who is potentially transgender. Researchers can then examine these sub-populations to determine if there is an adequate representation within the sample and to see if there may be reason to suspect that the efficacy and safety of the test drug is different for this sub-population. This solution can help clinicians determine if transgender subjects may have a different response to the test drug.}

Another consideration when incorporating gender or sex information into a statistical model is to be aware of the potential biases from imputing missing data. There are many programs that attempt to infer a person's gender using data mining techniques. The possible biases resulting from these methods depend on the basis with which these packages determine gender and on whether gender identity or sex at birth (or something else) is the information being imputed. The gender R package for example, categorizes a person as ``male'' or ``female'' based historical data of gendered names~\citep{Mullen2018}. Therefore, as acknowledged in the package documentation, in no way can this package determine a person's gender identity or can it determine any specific details about a person's sexual characteristics. Furthermore, it can only partially identify subjects from historically marginalized gender roles as it does not measure transgender or other gender-non-conforming individuals. Another tactic that is often used to impute subjects' missing gender or sex classifications is to use crowdsourcing techniques. Most often, these methods rely on people who are paid to look at names and sometimes pictures of the subjects and then make a personal judgment to classify the subject as male or female. Hence, this method is subject to both personal and cultural bias and prone to misclassification errors. 

Collecting data in a way that honors different gender identities will contribute to an increase in the general knowledge of and respect for various gender identities in the greater scientific community. The results of these efforts could have an enormous effect on the quality of life of gender and sex minorities both by setting better scientific standards for the ways in which human subjects are treated with respect and by providing better targeted health care through more representative data analyses. Since knowledge of gender and sex demographics can better inform a clinician in their own practice, the statistical argument for more informative data and the usefulness of the resulting statistical models is highly connected to the ethical argument for inclusive language and data collection procedures. 
    
\section{Statistical considerations and practical guidelines}
Here we present a three-step approach for statistical practitioners to consider whenever collecting data on human sex and gender. Broadly speaking, the steps we encourage each researcher to take are to 1) identify the information relevant to the study; 2) respect the study participants' identities with inclusive language; 3) protect the data and participants' identities. We discuss statistical considerations that arise within these steps and offer some solutions for common issues that may arise, and apply these suggestions in two real examples.

\subsection{Identify relevant information}Gender and sex information does not necessarily need to be included in every study. Before researchers begin a project, it is important to determine whether it possible that gender or sex may affect the efficacy of the treatment. If researchers do not believe that gender or sex can moderate an effect, then they may not need to ask for this sensitive information. However, in many clinical settings in particular, the researcher may expect gender identity or sex to significantly affect the efficacy of a treatment. In such cases, it is important for the researcher to explicitly state the underlying mechanisms. For instance, if sex may affect the efficacy, which sexual characteristics specifically are hypothesized to affect the treatment? If someone is receiving hormone therapy treatment, is this an important factor in the efficacy of the treatment under study? If someone has undergone gender-affirming surgery, is this an important factor in the efficacy of the treatment? Ultimately, asking more specific questions about bodily attributes rather than using gender or sex as a catch-all variable will lead to more reliable and inclusive data. For example, cisgender women, transgender women, and transgender men may all need breast cancer screening. Therefore, a survey that asks patients for their gender identity only will exclude transgender men and conversely, a survey that asks patients for their sex assigned at birth only, will likely exclude transgender women. 

Often, researches use the words ``sex'' and ``gender'' as interchangeable characteristics. This is a dangerous practice in that it will most certainly lead to some type of misclassification bias. If gender or sex are relevant features of a statistical model, the researcher must dig deeper and ask is ``What is relevant to my study, gender or sex or both?'' This is merely an example of the more general statistical best practice, when collecting data from which to construct a statistical model, the researcher must carefully consider what are the variables of interest. Although there are many statistical techniques for variable selection, there is no equal substitute for domain-specific knowledge when it comes to variable selection. Example 2 provides an example of a clinical study where gender and sex information were conflated, influencing the conclusions of the data analysis. 

\textbf{Example 2}

\textbf{\textit{Setting: }} \textit{One of the authors (ST) served as a methodologist for a research project that developed a predictive model for a certain type of drug-resistant epilepsy. The researchers were interested in determining the efficacy of an anti-epileptic drug on people with different types of epilepsy~\citep{ChoiEtAl2020}. To predict whether or not this drug will reduce seizures in various types of patients, the researchers conducted a retrospective study and developed a predictive model based on data that had been collected many years prior across two different clinical groups. Useful predictive variables included age; different seizure types; whether or not there was a change in seizure frequency with a patient's menstrual cycle (called Catamenial epilepsy); and the sex of the patient (which was stated without explanation and only marked as male or female). }

\textbf{\textit{Problem:}} \textit{ The most important predictive variables are heavily related to one another. In particular, Catamenial epilepsy and sex contain overlapping information. The solution proposed by the statistician was to develop a new, three-level categorical predictor variable that indicated both sex and the occurrence of Catamenial epilepsy rather than to use each of these variables separately. Upon further reflection, the statistician started to realize the limitations of assessing signals from ambiguous data collection procedures regarding sex and gender and realized that the data collection procedure regarding the sex of the patients in this study was insufficient for capturing what appears to be the most relevant information about this variable. }

\textbf{\textit{Proposed solution:}} \textit{ For future studies, at the data collection stage, rather than asking for a patient's sex, more informative data could be easily obtained by instead asking patients both for their gender identity and whether or not they regularly experienced menses. This simple adjustment to the data collection stage would prevent the exclusion or missclassification of transgender men (who may still experience menses) and cisgender women who may not experience menses due to age or medical procedures (such as a hysterectomy) or having non-ovary producing internal genitalia. }

In a clinical trial setting where a group of many typical characteristics of human sex development is of interest, an inclusive question may ask participants for their sex assigned at birth. This will typically be either assigned male at birth (AMAB) or assigned female at birth (AFAB). However, it is worth considering how important information about intersex people may be lost in this classification system or how important information about transgender people may be lost if gender identity is not also specified. Furthermore, in the US and in other countries, gender neutral birth markers (X) are an increasingly popular option for new parents.

Research has shown that individuals from the LGBTQ+ population tend to experience higher rates of depression and suicidal ideation as well as perceived and experienced threats to personal safety~\citep{AlmeidaEtAl2009}. If mental health could possibly moderate the effect of a treatment, then sexual orientation and gender identity might be important variables to collect.

Researchers may be interested in gender identity or sexual characteristics of participants, but must ensure binary sex is not being relied upon as a proxy for another more precise variable such as the presence of breast tissue or type of hormonal balances. Imagine, for instance, designing a study to identify determinants of prostate cancer. Rather than ask patients for their sex, one may instead simply ask patients if they have (or have ever had) a prostate. This type of gender-neutral re-wording of survey questions is much more specific and can collect more relevant information in a way that respects individuals' identities. These measures can help avoid the confusion of dealing with transient variables such as sex or gender, both of which may change with time depending on how they are defined. Furthermore, this type of gender-neutral questioning avoids causing any unnecessary mental trauma to transgender and gender-non-conforming individuals who may feel outted or experience intense gender dysphoria as a result of being forced to identify with a binary gender that does not match their own identity. Figure 3 outlines general steps for evaluating what information is relevant for differently motivated studies. 

\begin{figure}
\includegraphics[width=\textwidth]{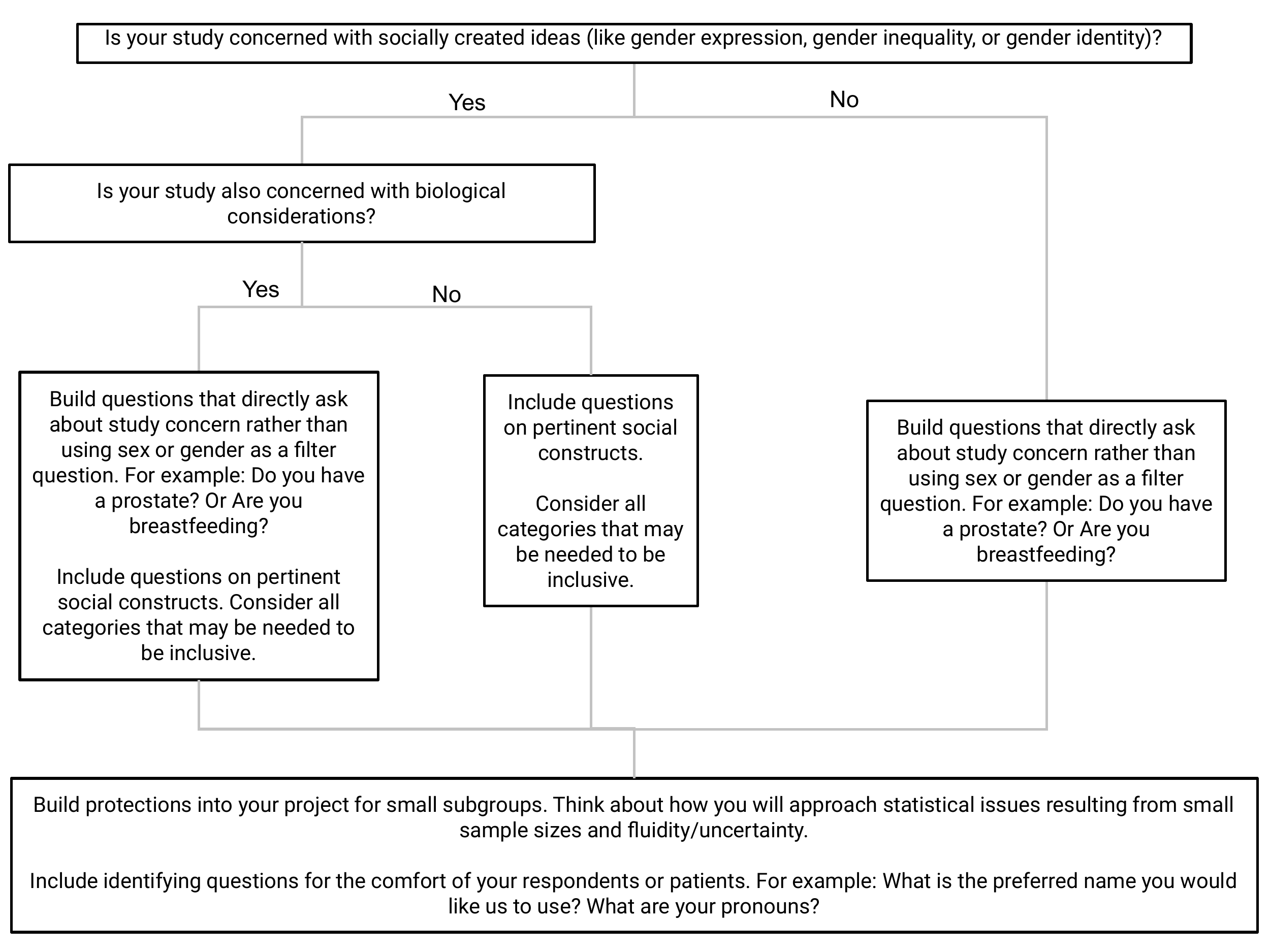}
\caption{A flow-chart outlining how to identify the relevant gender and sex information to reference in a particular study.}
\end{figure}

\subsection{Center inclusivity and respect }The purpose of institutional review boards is to protect the rights and welfare of human subjects. Human welfare is intimately connected with perceived respect, especially for those that identify with various socially and politically marginalized groups~\citep{Durwood2017}. The discretization of social constructs for the purpose of statistical analysis is not new and neither, unfortunately, are data collection processes that disregard individual identities. For example, it was not until 1960 that individuals could select their own racial identity in the US Census~\citep{Brown2020}. As a consequence of the National Research Act being signed into law in 1974, the Belmont Report identified ``respect for persons'' as a basic ethical principle of research. The report defines ``respect for persons'' based on two fundamental ethical convictions: first, that individuals should be treated as autonomous agents, and second, that persons with diminished autonomy are entitled to protection~\citep{Belmont}. It is critical for the researcher to take extra measures to ensure the respectful treatment of human subjects is an intrinsic feature of any human-centric scientific endeavor. Unfortunately, the considerations necessary for the respectful treatment of gender and sex minorities are too often overlooked.

One important measure that expresses respect to participants is the explicit communication of the purpose behind collecting sensitive data. Participants have the right to understand why their information is being collected and how it is going to be used and their identities protected. As discussed in the previous section, another important consideration is to recognize that questions about sex and gender may cause an undue amount of stress or confusion for transgender and gender-non-conforming respondents. If it is necessary to question a study participant about their sex, the researcher should clearly define what is meant by ``sex'' in a way that does not coerce the participant to adopt an inappropriate gender identity but instead emphasized specific sexual characteristics without gendered language. For example, in many non-clinical studies, gender is recorded because the variable of interest is whether or not an individual belongs to a historically marginalized group. In such cases, it would be misleading to discount the marginalization of transgender men, say, by grouping them together with cisgender men. In this case, the question about gender could be reworded to instead ask participants if they are cisgender men, cisgender women, transgender (men or women) or otherwise gender-non-conforming. As another example, if the outer genitalia a person currently has is a key feature of interest, it's best to ask this explicitly rather than ask the participant if they are male or female. On the other hand, if the genitalia a person was born with is the key factor, then the question can be easily adapted to adjust for this without challenging the participant's gender identity. 

In situations where the data analyst has no control over the data collection procedure, such as in retroactive studies, the analyst is responsible for noting any ambiguities or possible sources of bias in the data at hand. If gender information was collected but there were no options for people who did not identify as a man or a woman, this is important to note because the study is limited by not applying to gender-non-comforming populations. If the retroactive study collected information on the sex of subjects, the data analyst should strive to understand this variable in the context of the original study. In the resulting analysis, the researcher can still demonstrate respect for the human subjects by including a discussion on how sex was coded and, following the suggestions in the previous section, by discussing what particular sex characteristics were probably of interest using gender-neutral language. The importance of this careful and thoughtful data collection is only growing as many states (and countries) begin to offer gender-neutral identifiers on official documentation such as birth certificates and drivers' licenses~\citep{Fitzsimons2019}.

Even if the data analyst can control the manner in which the data is collected, creating additional categories such as ``non-binary'' for gender or ``intersex'' for sex can create difficulties in the statistical analysis if small sample size considerations are not taken into account. Whether or not a particular category is a valid option for a variable in a statistical model often comes down to an issue of sample size. To avoid this small sample size problem that often results in wasted data, the researcher must take time before collecting the data to plan the study with carefully developed sampling strategies and/or with explicitly defined questions that take into account the topics discussed in the previous section. Obviously these measures are important for studies that focus on the LGBTQ+ community. These considerations are especially important for gender and sex minorities, c.f. e.g. ~\citet{LauEtAl2020}; ~\citet{GoldsteinEtAl2017}; and ~\citet{GuptaEtAl2016}. %Example 3 outlines recent clinical trials which were specifically designed to account for small sample sizes issues thus better ensuring the resulting treatment would be effective across all minority populations. 

%\textbf{Example 3}

%\textbf{\textit{Setting:}} \textit{ The vaccine trials for COVID-19 undertaken by Pfizer and Moderna Inc in 2020 were designed to prove the safety and efficacy of the newly developed vaccine across the vast population ~\citet{1017470:21550441}. Inferences based on these trials would affect decisions and policy changes to the distribution of the vaccine and influence out understanding of the population at large and therefore must be applicable to the entire population, including minority groups. }

%\textbf{\textit{Possible problem:}} \textit{ Collecting a representative sample is especially important when the information from a statistical analysis may have serious consequences that impact the safety and health of a group of people. For underrepresented minority sub-populations, we may not truly understand the prevalence of a disease or the safety or efficacy of a treatment if this sub-population is not represented in our sample.}

%\textbf{\textit{Solution: }} \textit{Both companies designed phase three of their clinical trials to ensure that different minority demographics were adequately represented of the American population. Event thought the cost was an extension in the timeline of the development of the vaccines, both of these trials took extra care to increase the sample size based on an interim review of the demographic distribution of the recruited subjects~\citet{1017470:21550459}.}

Some additional considerations for surveys that take into account the points discussed above. When asking about participants' sex, consider including intersex as an option or if possible, consider rewording the question to ask about the specific sexual characteristics of interest. When asking about gender identity for classification purposes, include at least a third option for non-binary or gender non-conforming individuals. Similarly, if asking for participant pronouns be sure to include a gender neutral option such as the singular ``they'' in addition to the typical binary options of she/her/hers and he/him/is. When it is important to identify transgender and other gender and sex minority groups, we recommend using the two-step approach, asking participants both for their sex assigned at birth and for their gender identity~\citep{Badgett2009}.

\subsection{Protect the participant and the data}In our day-to-day activities, we regularly find ourselves being asked to provide our sex for events such as doctor's appointments, filling out the Census, registering a profile on a social media page, or applying for a job. For cisgender people, these questions may not cause a second through but for transgender and gender-non-conforming people, these questions can elicit not just mild discomfort but feelings of dread and fear for personal safety. There are many reasons why a person may not want to publicly reveal that their sex assigned at birth does not align with their gender identity; c.f. e.g., ~\citet{SaferEtAl2016} and ~\citet{WirtzEtAl2020}. From a statistical perspective, is important to recognize these concerns from the perspective of gender and sex minorities. If data protection and personal privacy cannot be guaranteed to these groups where the degree of potential harm from data misuse is great, there may be a severe non-response bias. More importantly, an individual's safety may be under threat. Of course from an ethical perspective, scientists recognize that data collection should not cause undue levels of stress to participants~\citep{BennounaEtAl2017}. There are two levels of data security to consider when collecting information on a person's gender identity or sex. The security of this information is crucial when collecting the data and when analyzing or sharing the data. 

In practice, data privacy is complicated by a lack of general or consistent standards. Differing approached to privacy can be illustrated by considering the parts of speech used. The usual method for protecting data privacy is through nouns: creating a list of data objects that qualify for protection as Personally Identifying information (PII). This results in conflicting standards as each regulating organization may have a different list. For example, the US Department of Homeland Security has a list of nineteen items, six of which are considered PII alone and nine are considered PII if paired with another item on the list~\citep{DHS2012}. However, the US Department of Health and Human Services has a single tier of 18 items. Only thirteen items are on both lists. Neither protects gender; sexual orientation qualifies as Sensitive PII for Homeland Security but is left unprotected by HHS. Modern analytic algorithms pose an additional problem: identifying persons through a multi-point match of data objects appearing on no list. A different approach to data privacy is suggested Australian Data Privacy act, established in 2008 and updated it in 2014 and 2019~\citep{MorrisCravigan2020}. This act takes a verb-based approach to privacy, placing restrictions where an individual ``reasonably identifiable''. Taking this notion a step further, we can better understand data privacy as an adverb: the manner in which things are done. Assuring data, personal identify, and privacy are protected with adverbs - securely, transparently, ethically - achieves the desired intent. 

Socially typical safety standards may not be adequate especially for transgender individuals~\citep{RoffeeWaling2017}. As noted, the US HHS standard conflicts with others and affords no protection to sexual identity data. The unfortunate reality is that many gender-non-conforming children may not feel safe to reveal their gender identities if a parent or legal guardian is present; c.f. e.g. ~\citet{RiggsBartholomaeus2018} and ~\citet{CatalpaMcGuire2018}. Although we are limiting our discussion in this article to data collection for adults, note that extra considerations are needed if the participants of a study are minors. These considerations for safety of the subject participants are relevant for adults as well. If a person is verbally asked to identify their sex and/or gender identity in say a doctor's waiting room, a transgender patient may feel vulnerable to harassment to others in the room. Thus, it is important that security and privacy measures are taken into account at the data collection stage by minimizing the opportunities for discrimination and harassment of the participants. 

The privacy of personal information, especially in small studies, can leave minorities vulnerable to discovery if the information collected is too specific. For example, a transgender patient's identity may be identifiable from the data if it includes particular hospital information or information about a specific type of treatment. It is possible that gender and sex minorities may be identified even if a study adheres to HIPPA requirements~\citep{Corliss2015}. While a full discussion of data identifiability concerns is beyond the scope of this article, a tried-and-true statistical recommendation is to plan studies to avoid small sample size problems. This may require budgeting time or money to implement carefully designed sampling procedures. According to a 2018 Gallup poll, approximately 5\% of the US adult population identify as LGBTQ+. Because gender and sex minorities represent a small proportion of the overall population, any survey collecting gender or sex data is likely to face the issue of unavoidably small sample sizes. In such instances, we recommend analyzing aggregated information rather than rely on a single survey for analysis. There are many statistical methods for conducting meta-analyses on aggregated data and the benefits of such methods may include higher statistical power in addition to stronger data privacy~\citep{ChaudhuriWeinberg2017,Gatewood2001,Abowd2018}. 
    
\section{Discussion}
LGBTQ+ individuals have always been part of society, even if these identities have not been encoded in previous data collection efforts. Much research concerns response variables that can be moderated by gender, sexual characteristics, or both. To prevent the continued erasure of minority populations from scientific research, it is time for the standards of data collection procedures to include options outside of the traditional gender and sex binary. Another important result of refining how gender and sex data is collected will be a decrease in measurement error and bias in the results of statistical analyses.
\appendix 

\section{Appendix}\label{appendix-title-8a6d1a48f568}
    For the interested reader, we have compiled some additional material that may be of use for implementing more inclusive data collection procedures. Example of inclusive demographic survey questions can be found on many survey sites such as Alchemer \BreakURLText{(https://www.alchemer.com/resources/blog/how-to-write-better-demographic-questions).} There are also many consulting and training resources offered by companies such as Speaking of Transgender \BreakURLText{(https://speakingoftransgender.com/transgender-training-and-course)} and Sylveon Consulting \BreakURLText{(https://www.sylveonconsulting.com).} Writing tools and guidance are provided by many freelance professionals and professional groups such as transgender author Julia Serano \BreakURLText{(https://www.patreon.com/juliaserano/overview),} The Society of Professional Journalists \BreakURLText{(https://www.spj.org/race-gender-hotline.asp),} Radical Copy Editor\\ \BreakURLText{(https://radicalcopyeditor.com/2017/08/31/transgender-style-guide)}

A wealth of additional educational material can be found on the websites for the Federal Committee on Statistical Methodology \BreakURLText{(https://nces.ed.gov/FCSM/SOGI.asp),} Planned Parenthood \BreakURLText{(https://www.plannedparenthood.org/learn/gender-identity),} The LGBT Foundation\\ \BreakURLText{(https://lgbt.foundation/research-guidance),} and The National LGBTQIA+ Health Education Center \BreakURLText{(https://www.lgbthealtheducation.org/resources/type/publication).}

\bibliographystyle{plainnat}

\bibliography{Best_Practices_for_Collecting_Gender_and_Sex_Data}

\begin{thebibliography}{42}
\providecommand{\natexlab}[1]{#1}
\providecommand{\url}[1]{\texttt{#1}}
\expandafter\ifx\csname urlstyle\endcsname\relax
  \providecommand{\doi}[1]{doi: #1}\else
  \providecommand{\doi}{doi: \begingroup \urlstyle{rm}\Url}\fi

\bibitem[Abowd(2018)]{Abowd2018}
J.~M. Abowd.
\newblock The us census bureau adopts differential privacy.
\newblock \emph{Proceedings of the 24th ACM SIGKDD International Conference on
  Knowledge Discovery \& Data Mining}, pages 2867--2867, 2018.

\bibitem[Ainsworth(2018)]{Ainsworth2018}
C.~Ainsworth.
\newblock Sex redefined: The idea of 2 sexes is overly simplistic.
\newblock \emph{Scientific American}, 22, 2018.

\bibitem[Almeida et~al.(2009)Almeida, Johnson, Corliss, Molnar, and
  Azrael]{AlmeidaEtAl2009}
J.~Almeida, R.~M. Johnson, H.~L. Corliss, B.~E. Molnar, and D.~Azrael.
\newblock Emotional distress among lgbt youth: The influence of perceived
  discrimination based on sexual orientation.
\newblock \emph{Journal of youth and adolescence}, 38:\penalty0 1001--1014,
  2009.

\bibitem[Alzahrani et~al.(2019)Alzahrani, Nguyen, Ryan, Dwairy, McCaffrey,
  Yunus, Forgione, Krepp, Nagy, Mazhari, and Reiner]{AlzahraniEtAl2019}
T.~Alzahrani, T.~Nguyen, A.~Ryan, A.~Dwairy, J.~McCaffrey, R.~Yunus,
  J.~Forgione, J.~Krepp, C.~Nagy, R.~Mazhari, and J.~Reiner.
\newblock Cardiovascular disease risk factors and myocardial infarction in the
  transgender population.
\newblock \emph{Circulation: Cardiovascular Quality and Outcomes}, 12\penalty0
  (4), 2019.

\bibitem[Association(2018)]{ASAethics}
American~Statistical Association.
\newblock Ethical guidelines for statistical practice.
\newblock Available at
  \url{https://www.amstat.org/asa/files/pdfs/EthicalGuidelines.pdf}, 2018.
\newblock Committee on Professional Ethics.

\bibitem[Badgett and Team(2009)]{Badgett2009}
M.~Badgett and Sexual Minority Assessment~Research Team.
\newblock Best practices for asking questions about sexual orientation on
  surveys.
\newblock \emph{UCLA: The Williams Institute}, 2009.

\bibitem[Bates et~al.(2019)Bates, Trejo, and Vines]{BatesEtAl2019}
N.~Bates, Y.~A.~García Trejo, and M.~Vines.
\newblock Are sexual minorities hard-to-survey? insights from the 2020 census
  barriers, attitudes, and motivators study (cbams) survey.
\newblock \emph{Journal of Official Statistics}, 35\penalty0 (4):\penalty0
  709--729, 2019.

\bibitem[Bauer et~al.(2019)Bauer, Devor, Heinz, Marshall, Sansfaçon, and
  Pyne]{BauerEtAl2019}
G.~Bauer, A.~Devor, M.~Heinz, Z.~Marshall, A.~Pullen Sansfaçon, and J.~Pyne.
\newblock Cpath ethical guidelines for research involving transgender people \&
  communities.
\newblock \emph{Canadian Professional Association for Transgender Health},
  2019.

\bibitem[Bennouna et~al.(2017)Bennouna, Mansourian, and
  Stark]{BennounaEtAl2017}
C.~Bennouna, H.~Mansourian, and L.~Stark.
\newblock Ethical considerations for children’s participation in data
  collection activities during humanitarian emergencies: A delphi review.
\newblock \emph{Conflict and Health}, 11\penalty0 (1):\penalty0 1--15, 2017.

\bibitem[Brown(2020)]{Brown2020}
A.~Brown.
\newblock The changing categories the u.s. census has used to measure race.
\newblock Pew Research Center, 2020.

\bibitem[Catalpa and McGuire(2018)]{CatalpaMcGuire2018}
J.~M. Catalpa and J.~K. McGuire.
\newblock Family boundary ambiguity among transgender youth.
\newblock \emph{Family Relations}, 67\penalty0 (1):\penalty0 88--103, 2018.

\bibitem[Choi et~al.(2020)Choi, Detyniecki, Bazil, Thornton, Crosta, Tolba, and
  et~al.]{ChoiEtAl2020}
H.~Choi, K.~Detyniecki, C.~Bazil, S.~Thornton, P.~Crosta, H.~Tolba, and
  M.~Muneeb et~al.
\newblock Development and validation of a predictive model of drug-resistant
  genetic generalized epilepsy.
\newblock \emph{Neurology}, 95\penalty0 (15):\penalty0 e2150--e2160, 2020.

\bibitem[Clements-Nolle et~al.(2006)Clements-Nolle, Marx, and
  Katz]{ClementsNolle2006}
K.~Clements-Nolle, R.~Marx, and M.~Katz.
\newblock Attempted suicide among transgender persons: The influence of
  gender-based discrimination and victimization.
\newblock \emph{Journal of Homosexuality}, 51\penalty0 (3):\penalty0 53--69,
  2006.

\bibitem[Corliss(2015)]{Corliss2015}
D.~J. Corliss.
\newblock Analytics and protected healthcare information.
\newblock Presentation for Michigan Association of Community Mental Health
  Boards \url{https://www.peace-work.org/peace-work-studies}, 2015.

\bibitem[DHS(2012)]{DHS2012}
US~DHS.
\newblock Handbook for safeguarding sensitive personally identifiable
  information, 2012.

\bibitem[Dubois(2012)]{Dubois2012}
L.~Z. Dubois.
\newblock Associations between transition-specific stress experience, nocturnal
  decline in ambulatory blood pressure, and c-reactive protein levels among
  transgender men.
\newblock \emph{American Journal of Human Biology}, 24\penalty0 (1):\penalty0
  52--61, 2012.

\bibitem[Durwood et~al.(2017)Durwood, McLaughlin, and Olson]{Durwood2017}
L.~Durwood, K.~A. McLaughlin, and K.~R. Olson.
\newblock Mental health and self-worth in socially transitioned transgender
  youth.
\newblock \emph{Journal of the American Academy of Child and Adolescent
  Psychiatry}, 56\penalty0 (2):\penalty0 116--123, 2017.

\bibitem[Fausto-Sterling(2000)]{FaustoSterling2000}
A.~Fausto-Sterling.
\newblock \emph{Sexing the body: Gender politics and the construction of
  sexuality}.
\newblock Basic Books, 2000.

\bibitem[Fitzsimons(2019)]{Fitzsimons2019}
T.~Fitzsimons.
\newblock N.j. to become fourth state with gender-neutral birth certificate
  option.
\newblock NBC News, 2019.

\bibitem[Gatewood et~al.(2001)]{Gatewood2001}
George Gatewood et~al.
\newblock A monograph on confidentiality and privacy in the us census.
\newblock \emph{US Bureau of the Census, Washington}, 2001.

\bibitem[Goldstein et~al.(2017)Goldstein, Corneil, and
  Greene]{GoldsteinEtAl2017}
Z.~Goldstein, T.~A. Corneil, and D.~N. Greene.
\newblock When gender identity doesn't equal sex recorded at birth: the role of
  the laboratory in providing effective healthcare to the transgender
  community.
\newblock \emph{Clinical chemistry}, 63\penalty0 (8):\penalty0 1342--1352,
  2017.

\bibitem[Greenberg and Stam(2012)]{GreenbergStam2012}
J.~A. Greenberg and R.~Stam.
\newblock \emph{Intersexuality and the Law : Why Sex Matters}.
\newblock 2012.

\bibitem[Gupta et~al.(2016)Gupta, Imborek, and Krasowski]{GuptaEtAl2016}
S.~Gupta, K.~L. Imborek, and M.~D. Krasowski.
\newblock Challenges in transgender healthcare: the pathology perspective.
\newblock \emph{Laboratory medicine}, 47\penalty0 (3):\penalty0 180--188, 2016.

\bibitem[Herdt(2020)]{Herdt2020}
G.~Herdt.
\newblock \emph{Third sex, third gender: Beyond sexual dimorphism in culture
  and history}.
\newblock Princeton University Press, 2020.

\bibitem[HEW(1979)]{Belmont}
HEW.
\newblock The belmont report: Ethical principles and guidelines for the
  protection of human subjects of research.
\newblock Available at
  \url{http://www.nuhresearch.nhg.com.sg/obr/Belmont\%20Report.pdf}, 1979.

\bibitem[Hughes et~al.(2006)Hughes, Houk, Ahmed, Lee, and for
  Paediatric~Endocrinology]{HughesEtAl2006}
I.~A. Hughes, C.~Houk, S.~F. Ahmed, P.~A. Lee, and Lawson Wilkins Pediatric
  Endocrine Society/European~Society for Paediatric~Endocrinology.
\newblock Consensus statement on management of intersex disorders.
\newblock \emph{Journal of pediatric urology}, 2\penalty0 (3):\penalty0
  148--162, 2006.

\bibitem[interACT(2020)]{interACT}
interACT.
\newblock Statement on intersex terminology.
\newblock Available at
  \url{https://interactadvocates.org/interact-statement-on-intersex-terminology/},
  2020.

\bibitem[Irwig(2018)]{Irwig2018}
M.~S. Irwig.
\newblock Cardiovascular health in transgender people.
\newblock \emph{Reviews in Endocrine and Metabolic Disorders}, 19\penalty0
  (3):\penalty0 243--251, 2018.

\bibitem[Lau et~al.(2020)Lau, Antonio, Davison, Queen, and Devor]{LauEtAl2020}
F.~Lau, M.~Antonio, K.~Davison, R.~Queen, and A.~Devor.
\newblock A rapid review of gender, sex, and sexual orientation documentation
  in electronic health records.
\newblock \emph{Journal of the American Medical Informatics Association},
  27\penalty0 (11):\penalty0 1174--1783, 2020.

\bibitem[Morris and Cravigan(2020)]{MorrisCravigan2020}
M.~Morris and E.~Cravigan.
\newblock The privacy, data protection and cybersecurity law review: Australia.
\newblock
  \url{https://thelawreviews.co.uk/title/the-privacy-data-protection-and-cybersecurity-law-review/australia},
  2020.

\bibitem[Mullen(2018)]{Mullen2018}
L.~Mullen.
\newblock gender: Predict gender from names using historical data.
\newblock Available at \url{https://github.com/ropensci/gender}, 2018.
\newblock R package version 0.5.2.

\bibitem[Olson and McLaughlin(2016)]{OlsonMCLaughlin2016}
K.~R. Olson and K.~A. McLaughlin.
\newblock Mental health of transgender children who are supported in their
  identities.
\newblock \emph{Pediatrics}, 137\penalty0 (3), 2016.

\bibitem[Onyiriuka et~al.(2016)Onyiriuka, Kuhnle-Krahl, Sadoh, and
  Elusiyan]{Onyiriuka2016}
A.~N. Onyiriuka, U.~Kuhnle-Krahl, E.~Sadoh, and J.~B.~E. Elusiyan.
\newblock External genital anomalies in newborns in two nigerian hospitals: A
  pilot study of the birth prevalence.
\newblock \emph{International Journal of Child and Adolescent Health},
  9\penalty0 (2):\penalty0 187--193, 2016.

\bibitem[Pascoe and Richman(2009)]{PascoeRichman2009}
E.~A. Pascoe and L.~Smart Richman.
\newblock Perceived discrimination and health: A meta-analytic review.
\newblock \emph{Psychological Bulletin}, 135\penalty0 (4):\penalty0 531--554,
  2009.

\bibitem[Riggs and Bartholomaeus(2018)]{RiggsBartholomaeus2018}
D.~W. Riggs and C.~Bartholomaeus.
\newblock Gaslighting in the context of clinical interactions with parents of
  transgender children.
\newblock \emph{Sexual and Relationship Therapy}, 33\penalty0 (4):\penalty0
  382--394, 2018.

\bibitem[Roffee and Waling(2017)]{RoffeeWaling2017}
J.~A. Roffee and A.~Waling.
\newblock Resolving ethical challenges when researching with minority and
  vulnerable populations: Lgbtiq victims of violence, harassment and bullying.
\newblock \emph{Research Ethics}, 13\penalty0 (1):\penalty0 4--22, 2017.

\bibitem[Russell et~al.(2018)Russell, Pollitt, Li, and
  Grossman]{RussellEtAl2018}
S.~T. Russell, A.~M. Pollitt, G.~Li, and A.~H. Grossman.
\newblock Chosen name use is linked to reduced depressive symptoms, suicidal
  ideation, and suicidal behavior among transgender youth.
\newblock \emph{The Journal of adolescent health : official publication of the
  Society for Adolescent Medicine}, 63\penalty0 (4):\penalty0 503--505, 2018.

\bibitem[Safer et~al.(2016)Safer, Coleman, Feldman, Garofalo, Hembree, Radix,
  and Sevelius]{SaferEtAl2016}
J.~D. Safer, E.~Coleman, J.~Feldman, R.~Garofalo, W.~Hembree, A.~Radix, and
  J.~Sevelius.
\newblock Barriers to health care for transgender individuals.
\newblock \emph{Current opinion in endocrinology, diabetes, and obesity},
  23\penalty0 (2):\penalty0 168--168, 2016.

\bibitem[Saha-Chaudhuri and Weinberg(2017)]{ChaudhuriWeinberg2017}
P.~Saha-Chaudhuri and C.~R. Weinberg.
\newblock Addressing data privacy in matched studies via virtual pooling.
\newblock \emph{BMC Medical Research Methodology}, 17\penalty0 (1), 2017.

\bibitem[Su et~al.(2016)Su, Irwin, Fisher, Ramos, Kelley, Mendoza, and
  Coleman]{SuEtAl2016}
D.~Su, J.~A. Irwin, C.~Fisher, A.~Ramos, M.~Kelley, D.~A.~Rogel Mendoza, and
  J.~D. Coleman.
\newblock Mental health disparities within the lgbt population: A comparison
  between transgender and nontransgender individuals.
\newblock \emph{Transgender Health}, 1\penalty0 (1):\penalty0 12--20, 2016.

\bibitem[Wiesner-Hanks(2011)]{WiesnerHanks2011}
M.~E. Wiesner-Hanks.
\newblock \emph{Gender in history: Global perspectives}.
\newblock John Wiley \& Sons, 2011.

\bibitem[Wirtz et~al.(2020)Wirtz, Proteat, Malik, and Glass]{WirtzEtAl2020}
A.~L. Wirtz, T.~C. Proteat, M.~Malik, and N.~Glass.
\newblock Gender-based violence against transgender people in the united
  states: A call for research and programming.
\newblock \emph{Trauma, Violence, \& Abuse}, 21\penalty0 (2):\penalty0
  227--241, 2020.

\end{thebibliography}
\end{document}